\documentstyle[prl,floats,epsfig,aps]{revtex}
\input{psfig.sty}
\begin{document}

\title{Rigorous decimation-based construction of ground pure states \\
for spin glass models on random lattices.}

\vskip .5cm
\author{S. Cocco$^{1}$, O. Dubois$^{2}$, J. Mandler$^{2}$ and 
R. Monasson $^{3,4}$}
\address{$^1$  CNRS-Laboratoire de Dynamique des Fluides Complexes,
3 rue de l'Universit{\'e}, 67000 Strasbourg, France;\\
$^2$ CNRS-Laboratoire d'Informatique de Paris 6, 4 place Jussieu,
75005 Paris, France;\\ 
$^{3}$ CNRS-Laboratoire de Physique Th{\'e}orique de l'ENS,
24 rue Lhomond, 75005 Paris, France;\\
$^{4}$ CNRS-Laboratoire de Physique Th{\'e}orique,
3 rue de l'Universit{\'e}, 67000 Strasbourg, France.}
\date{\today}
\maketitle  

\begin{abstract}
A constructive scheme for determining pure states (clusters) at very
low temperature in the $3$-spins glass model on a random lattice is
provided, in full agreement with Parisi's one step replica symmetry 
breaking (RSB) scheme. Proof is based on the analysis of an exact 
decimation procedure. When the number $c$ of couplings per spin is 
smaller than some critical value $c_d$, all spins are eliminated
at the end of decimation (RS phase). In the range $c_d<c<c_s,$ a 
reduced Hamiltonian is left; each ground state (GS) of the latter is a
"seed" from which a cluster of GS of the original Hamiltonian can
be reconstructed. Above $c_s$, GS are frustrated
with an energy per spin larger than $-c$.
The number of GS in each cluster, the
number of clusters, the distances between GS are calculated and 
correspond to RSB predictions. 
\end{abstract}
\vskip .5cm

Parisi's replica symmetry breaking (RSB) theory and its physical
interpretation have turned out to be very fruitful for the 
investigation of disordered systems over the last twenty
years\cite{Mpv}. Unfortunately, the mathematical basis for RSB is
rather weak.  Rigorous studies confirming RSB predictions have been
limited to few mean-field models 
so far\cite{Talagrand}. One of the most striking assumptions
in RSB theory, the existence of numerous pure states, or clusters, in
phase space on which Gibbs measure becomes concentrated at low
temperature is still not fully understood even at the mean field level
despite recent progresses\cite{Talagrand}, not to speak about its
applicability to finite dimensional systems.

In this Letter, we present a rigorous study of a spin glass model
which allows us to identify explicitely pure states
for a given sample. Our analysis consists in decimating well chosen
spins appearing in the original Hamiltonian $H$. When decimation
stops, two cases may occur depending on the value of control
parameters.  If no spin is left, the partition function is entirely
known, as well as the properties of $H$; this situation corresponds
to the existence of a single pure state and replica symmetry
(RS). Otherwise, some reduced Hamiltonian $H'$ involving a subset
${\cal S}'$ of the original set of spins ${\cal S}$ has to be
treated. The ground states (GS) of $H'$ can be interpreted as 
seeds for the (low temperature) pure states of $H$. More precisely, all GS
in a pure state of $H$ can be reconstructed when backtracking 
the original decimation procedure from the seed of this cluster.  
Analysis of the statistical properties of seeds in the ${\cal S}'$ space, 
and of the reconstruction
process provides a full characterization of the structure of GS
in the ${\cal S}$ space.

The model we consider is the so-called 3-spins Ising spin glass. $M$ triplets 
of distinct integers ${i_m}< {j_m}< {k_m}$ are randomly chosen in the range 
$1,...,N$; plaquette $m$, formed by the attached spins, is associated 
a coupling $J_m$ equal to $\pm 1$ with equal
probabilities. The Hamiltonian
\begin{equation} \label{3sp}
H = - \sum _{m=1} ^M J_m \, S_{i_m}\, S_ {j_m} \, S_ {k_m} \ .
\end{equation}
equals $-M$ plus twice the number of frustrated plaquettes. We shall use $c$
to denote the number of plaquettes per spin, $M/N$. The thermodynamical 
properties of model (\ref{3sp}) were first investigated when all spins
interact together {\em i.e.}  in the limit of large ratios 
$c$\cite{Gardner,Td3sp}. Three phases were found. When the temperature $T$ is 
larger than $T_d(c) \sim 0.68 \sqrt c$, the system is paramagnetic 
(RS phase)\cite{Td3sp}. At low temperatures $T<T_s(c) \sim 0.65
\sqrt c$\cite{Gardner}, the system is trapped in one of the few existing 
glassy pure states. In the intermediate range $T_s<T<T_d$, there exist an 
exponential number of glassy pure states, separated by infinite 
barriers\cite{Td3sp}. The onset of ergodicity breaking at $T_d$, well above 
the equilibrium transition taking place at $T_s$, has made Hamiltonian 
(\ref{3sp}) a sensible mean field model of structural glass, 
and led to intense investigations of its out-of-equilibrium dynamical 
properties\cite{Td3sp,Glass}. 

As the ratio $c$ of plaquettes (interactions) per spin decreases, so do
the temperatures $T_d$ and $T_s$. The ratios $c_d \simeq 0.818$
and $c_s \simeq 0.918$ at which they respectively vanish were recently
calculated in the framework of one step RSB theory\cite{Rsb}, with the
following zero temperature picture\cite{Varsat}. For $c<c_s$
(respectively $c>c_s)$, the ground state (GS) of Hamiltonian
(\ref{3sp}) are unfrustrated (resp. frustrated) with an energy per
spin equal to (resp. larger than) $-c$. In the unfrustrated phase, the
number of GS scales as $2^{Ns}$ where the zero temperature entropy
simply equals $s=1-c$ (base 2 logarithm) (Fig.~1). With high
probability, two GS differ by a number of spins equal to $N d$ with
$d=1/2$. The spatial organization of GS in the space of configurations
${\cal S}$ ($N$-dimensional hypercube) undergoes a drastic change at
$c_d$ (Fig.~1), reminiscent of the ergodicity breaking taking place at
$T_d(c)$. The set of GS breaks into a large number, $2^{Ns_0}$, of
clusters, each containing an exponential number, $2^{Ns_1}$, of
GS. Two GS belonging to different clusters lie apart at a Hamming
distance $d_0=d=1/2$ while, inside a cluster, the distance is smaller,
and equal to $d_1=(1-b)/2$. $b$, the largest root of
\begin{equation}
b = 1 - e ^{- 3 \,c\, b^2} \quad ,
\label{backbone}
\end{equation}
measures the size of the cluster backbone, {\em i.e.} the fraction of spins 
common to all GS in a cluster. The entropies of clusters, 
$s_0=b-3 c b^2 + 2 c b^3$, and GS in a cluster, $s_1=s-s_0$, have been
computed within the RSB framework\cite{Rsb}. The corresponding curves
are shown on Fig.~1. At $c_d$, the total entropy $s$ is analytic in $c$, 
while the order parameter $b$, the entropies $s_0$, $s_1$ undergo 
discontinuous (first order) jumps e.g. from $b_d^-=0$ to $b^+_d \simeq 0.71$.

We now sketch how the above results may found back rigorously. Full
proofs will be given in an extended publication in a mathematical
journal. The techniques used are borrowed from probability theory,
and the analysis of algorithms. Their use was suggested from the  
close relationship between Hamiltonian (\ref{3sp}) and the random 
3-XORSAT optimization problem\cite{Rsb,Creignou,Focs}.
Let us call $\ell$-spin a spin which appears in $\ell$
distinct plaquettes in (\ref{3sp}). Plaquettes containing
at least a 1-spin are never frustrated. Our
decimation procedure consists in a recursive elimination of these
plaquettes and attached 1-spins (Fig.~2), until no 1-spin is left\cite{Uc}. 
We define the numbers $N_\ell (T)$ of $\ell$--spins after $T$ steps 
of the decimation algorithm, {\em i.e.} once $T$ plaquettes have been 
removed, and their set ${\cal N} (T)= \{ N_\ell(T) , \ell \ge 0\}$. 
The variations of 
the $N_\ell$s during the $\left( T+1\right)^{th}$ step of the algorithm
are stochastic variables due to the randomness in (\ref{3sp}) and
the choice of the 1-spin to be removed, with conditional expectations
with respect to ${\cal N} (T)$ given by
\begin{equation}
\label{flow}
{\mathbf{E}}[N_\ell(T+1) - N_\ell (T) |{\cal N}(T)]
= 2 \,p_{\ell+1}(T) -2 \,p_{\ell}(T) +
 \delta_{\ell,0}-\delta_{\ell,1}  \quad ,
\end{equation}
where $\delta$ denotes the Kronecker function.  When a plaquette is
removed, a 1-spin disappears (-$\delta_{\ell,1}$ term in (\ref{flow})) to
become a 0--spin ($\delta_{\ell,0}$). The plaquette contains two other
spins. The number of occurrences $\ell$ of each of these two spins is
distributed with probability $p_\ell (T) = \ell\,N_\ell(T)
/3/(M-T)$, and is diminished by one once the plaquette is taken
away. For large sizes $N$, the densities $n_\ell=N_\ell/N$ of
$\ell$--spins becomes self--averaging, and evolve on a long time scale
of the order of $N$\cite{Auto}. 
Defining the reduced time $t=T/N$, the densities
obey a set of coupled differential equations which can be deduced from
(\ref{flow}),
\begin{equation}
\label{diff}
\frac {d n_\ell }{dt}= \frac{2 \left[ (\ell+1)\, n_{\ell+1}(t)-\ell\,
n_{\ell}(t) \right]} {3(c-t)} + \delta_{\ell,0}-\delta_{\ell,1} \ .
\end{equation}
Initially, densities are Poisson distributed: $n_\ell(0)=e^{-3c}\,
{(3c)}^{\ell}/\ell!$. Equation (\ref{diff}) may be solved, with the
result  
\begin{equation}
\label{sol}
n_1 (t)= 3\,c\, b(t)^2\;\left( e^{-3\,c\,b(t)^2}+b(t)-1 \right) \ ,
\end{equation}
where $b(t) \equiv (1-t/c)^{1/3}$, while $n_{\ell}(t)$ is given by a Poisson
distribution of parameter $3\,c\,b(t)^2$ for $\ell \ge 2$. The density
of 1--spins is showed on Fig.~3 for various initial plaquettes per spin
ratios $c$. The algorithm stops at the time $t^*$ for which $n_1$ 
vanishes, that is, when no 1-spin is left. From eqn. (\ref{sol}),
$b(t^*)$ coincides with $b$ defined from eqn. (\ref{backbone})\cite{Add}.

What does the reduced Hamiltonian $H'$ look like once the decimation
has stopped? For $c<c_d$, $t^*=c$ and no spin and plaquette is left.
The entropy $s$ of (unfrustrated) GS of $H$ can be computed
recursively. Each time a plaquette containing $v (\ge 1)$ 1-spins and
these $v$ vertices are removed (Fig.~2), the number of GS gets
divided by $2^{v-1}$, and the average entropy (base 2 logarithm) of
GS decreased by ${\mathbf{E}}[v-1 |{\cal N}( T)]= 2\, p_1(T)$. As no
spin is left when the algorithm stops, the final value for the
entropy vanishes, giving
\begin{equation} \label{sent}
s= \int_0^{t^*}\, dt\, \frac{2\,n_1(t)}{3\, (c-t)} + e^{-3c} \ ,
\end{equation}
where the last term comes from the contribution $n_0(0)$ of 0-spins.
Using  eqn. (\ref{sol}) and $t^*=c$, we find back $s=1-c$.  
When $c_d<c<c_s$, the decimation procedure stops at $t^* < c$, and has 
not succeeded in eliminating all plaquettes and spins. The remaining 
fraction of plaquettes per spin,
$c'=(c-t^*)/\sigma$ where $\sigma =\sum_{\ell\ge 2} n_\ell(t^*)$, is plotted
as a function of $c$ in Inset of Fig.~3. 
Each GS of $H'$ can be seen as a `seed' from which a cluster of GS of $H$ 
in the original configuration space can be reconstructed. To do so,
plaquettes which were eliminated during decimation are reintroduced,
one after the other, and the spins they contain are
assigned all possible values that leave the plaquettes unfrustrated. 
Combining any of these partial spin assignments
with (free) 0-spins assignments, all the GS in a cluster are obtained.
Repeating the argument leading to the calculation of the entropy
in the $c<c_d$ case, we find that the average entropy $s_1$ of GS in a 
cluster
is precisely given by the r.h.s. of eqn. (\ref{sent}), and agrees with 
the RSB prediction. 

To complete our description of clusters, some statistical knowledge
about their seeds is required. The number ${\cal U}'$ of unfrustrated
GS of $H'$ can be analyzed by means of the first and second moments 
method\cite{Moments}, giving respectively some upper and lower bound to the
probability Pr$({\cal U}'\ge 1)$ of existence of unfrustrated GS,
\begin{equation} \label{moment}
\frac{{\mathbf{E}}\left( {\cal U}'\right) ^{2}}{{\mathbf{E}}
\left( {\cal U}'^{2}\right) } \leq \Pr \left( {\cal U}'\ge 1
\right) \leq {\mathbf{E}}\left(  {\cal U}' \right) \ .
\end{equation}
The right inequality is a consequence of the Markov bound for
positive variables,
$\Pr \left({\cal U}' \geq a\right) \leq {\mathbf{E}}\left( {\cal U}'
\right) /a$ with
$a=1$, while the left inequality can be established from
the Cauchy-Schwarz inequality, ${\mathbf{E}}
\left( {\cal U}' . {\cal V}\right) ^{2}\leq {\mathbf{E}}\left( 
{\cal U}'^{2}\right) .{\mathbf{E}}\left({\cal V}^{2}\right)$, taking 
${\cal V}\equiv 1 - \delta _{{\cal U}',0}$.
As shown below, the lower and upper bounds
to the threshold $c'_s$ separating unfrustrated and frustrated phases
obtained from eqn. (\ref{moment}) coincide, which allows an
exact determination of $c'_s$.

In the limit of a large number $N'$ of non--decimated spins, the first
moment depends only on the numbers of spins and plaquettes:
${\mathbf{E}}( {\cal U}' )= 2^{N'\left( 1-c'\right)}$.  From
(\ref{moment}), we conclude that ${\cal U}'$ almost surely vanishes
when $c'> 1$. On the contrary, the second moment is affected by the
existence of contraints on the minimal number (two) of occurrences of
spins in $H'$.  Its computation requires a combinatorial analysis of
the number of Hamiltonians $H$, {\em i.e.} ways of choosing plaquettes
and couplings in (\ref{3sp}), having a given pair of configurations
for GS.  As this number depends only on the distance $d'$ between the
two configurations, $\mathbf{E}({\cal U'}^{2}) $ may be expressed as a
combinatorial sum involving level-2 generalized Stirling numbers of
the second kind {\em i.e.} the number of ways to partition objects
(spins in plaquettes) into subsets (spin indices) having each at least
two elements \cite{Focs}. Very general asymptotic estimates for these
have recently been found, which involve parameters implicitly defined
by transcendental saddle-point equations \cite{stirling}. In the
present case and for $c'<1$, our sum has just one dominant exponential
term, which is precisely $4^{N'( 1-c')},$ or the square of the first
moment. The non-exponential contributions are insufficient to modify
this picture. So, for $c'<1$ and $N'\to\infty$, the l.h.s. of 
(\ref{moment}) is asymptotically equal to unity, and GS are almost 
surely unfrustrated. The value $c_s$ of
$c$ giving $c'=1$ is found from the analysis of the algorithm (Inset
of Fig.~3) to be $\simeq 0.918$. In addition, the entropy $s'_0=1-c'$
of UGS allows to find back the RSB expression for the entropy of
clusters, $s_0=\sigma\, s'_0$.

The self-averageness of ${\cal U}'$, {\em i.e.} of
the partition function $Z'\simeq {\cal U}' e^{M'/T}$ in the low temperature
$T\to 0$ limit, constrasts with the (sample--to--sample)
fluctuations exhibited by ${\cal U}$. Indeed, a direct 
application of inequalities (\ref{moment}) to ${\cal U}$ only 
permits to derive upper ($c_s\le 1$) and lower ($c_s\ge 0.889$) 
bounds to the threshold\cite{Creignou}. 
Fluctuations of ${\cal U}$ thus essentially come from fluctuations in 
the numbers $N_0$ and $N_1$ of 0- and 1-spins
(whose plaquettes form the dangling ends of the graph on Fig.~2) 
removed by the decimation algorithm. This comes as no surprise since
variations of  $N_0$ and $N_1$ induce 
drastic changes on the number of GS e.g.
the presence of a 0-spin multiply the number of GS by two.
Conversely, in $H'$, spins appear at least twice and are more 
interconnected, giving rise to weaker fluctuations for ${\cal U}'$. 

The reconstruction process allows a complete 
characterization of GS, in terms of an extensive number of (possibly 
overlapping) blocks made of few spins, each block being  
allowed to flip as a whole from a GS to another. When $c<c_d$, 
with high probability, two randomly picked GS differ over a fraction $d=1/2$
of spins, but are connected through a sequence of $O(N)$ successive GS 
differing over $O(1)$ spins only. For $c_d<c<c_s$, flippable blocks are 
juxtaposed to a set of seed-dependent frozen spins. The largest
Hamming distance $d_1 ^{max} (>d_1)$ between two GS associated to the 
same seed can be shown to be lower than the smallest possible distance
$d_0 ^{min} (\le d_0=1/2$) between any two seeds, thus proving the clustering 
property (Fig.~1).

The above results may be extended to multi $p$-spins interactions with
$p\ne 3$, with results in agreement with replica theory: the $p\ge 4$
case is qualitatively similar to $p=3$; for $p=2$, $c_d=c_s=1/2$ both
coincide with the percolation threshold.  Another extension regards
finite temperature. For $c<c_d$, the decimation algorithm allows a
complete calculation of the free energy. Whether 1-spins are set to
frustrate, or unfrustrate the plaquette they belong to, the energy is
increased, or decreased by one.  The resulting free energy density
equals $f(T)= -T \ln 2 - c\, T \ln \cosh (1/T)$ in agreement with the
replica paramagnetic calculation\cite{Rsb}. The same expression for
$f$ is likely to be established for $c_d<c<c_s$ through an extension
of the above approach.
Investigation of the frustrated region, $c>c_s$, is currently under
way, and would ultimately permit a rigorous construction of the
continuous RSB phase. Our results also sheds some light on the
observed coincidence between the onset of RSB and the failure of a
leaf removal algorithm in the vertex covering of random
graphs\cite{Vc}. Finally, it would be interesting to see how our
spin decimation approach could be extended 
to other distributions of random lattices.

{\bf Acknowledgments.} 
We thank the organizers of the SAT2002 meeting in Cincinnati (May
2002) during which part of this work was done.  Partial financial
support was provided by the French Ministry of Research (ACI Jeunes 
Chercheurs).
While writing this work, we became aware of an independent work by
M. Mezard, F. Ricci-Tersenghi and R. Zecchina on similar issues
\cite{MRZ}.

\twocolumn

\begin{figure}
\begin{center}
\includegraphics[width=190pt,angle=-90]{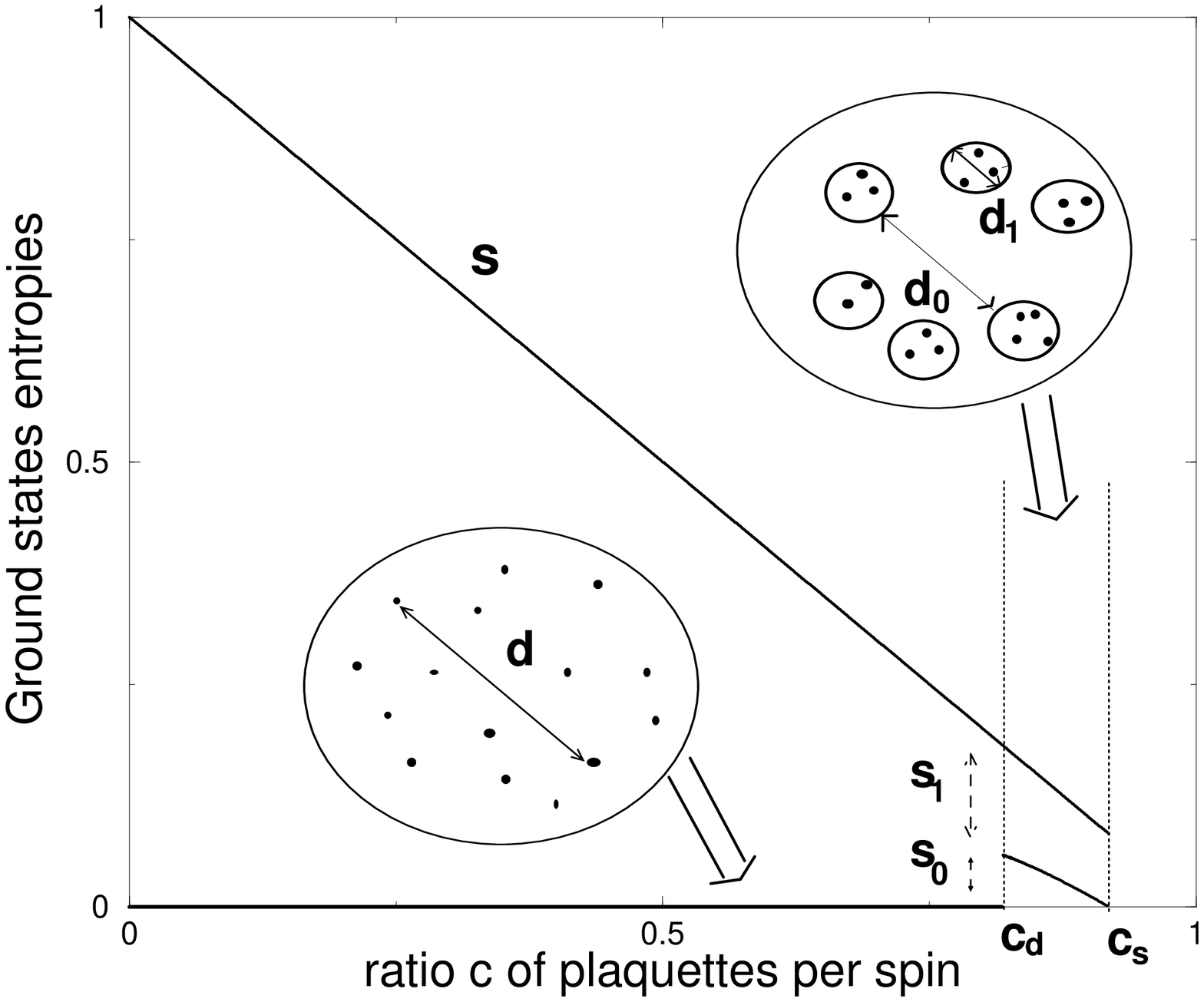}
\end{center}
\caption{GS structure and entropies as a function of the ratio $c$
of plaquettes per spin. The total entropy (log. number of unfrustrated
GS per spin) is $s=1-c$ for $c<c_s\simeq 0.918$. 
For $c<c_d\simeq 0.818$, GS are uniformely
scattered on the $N$-dimensional hypercube, with a typical normalized 
Hamming distance $d=1/2$.  
At $c_d$, the GS space discontinuously
breaks into disjoint clusters: the Hamming distance $d_1\simeq 0.14$ 
between solutions inside a cluster is much smaller than
the typical distance $d_0=1/2$ between two clusters
(RSB transition). The entropy of clusters, $s_0$,
and of solutions in each cluster, $s_1$, are such that $s_0+s_1=s$. 
At $c_s$, the number of clusters ceases to be exponentially large
($s_0=0$). Above $c_s$, GS are frustrated.}
\label{entro}
\end{figure}

\begin{figure}
\hskip .5cm
\begin{center}
\includegraphics[width=220pt,angle=-0]{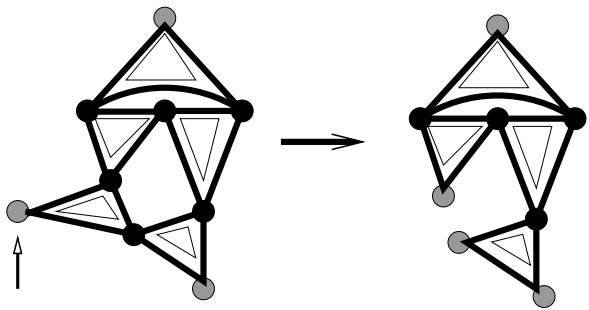}
\end{center}\caption{Graph representation of the 3-spins Hamiltonian.
Vertices (spins) are joined by plaquettes (values $\pm 1$ of couplings
are not shown here). A step of decimation consists in listing all
1-spins (gray vertices), choosing randomly one of them (gray vertex
pointed by the arrow), and eliminating this spin and its
plaquette. New 1-spins may appear. Decimation is repeated until no
1-spin is left.}

\label{plaqu}
\end{figure}

\begin{figure}
\begin{center}
\includegraphics[height=230pt,angle=-90]{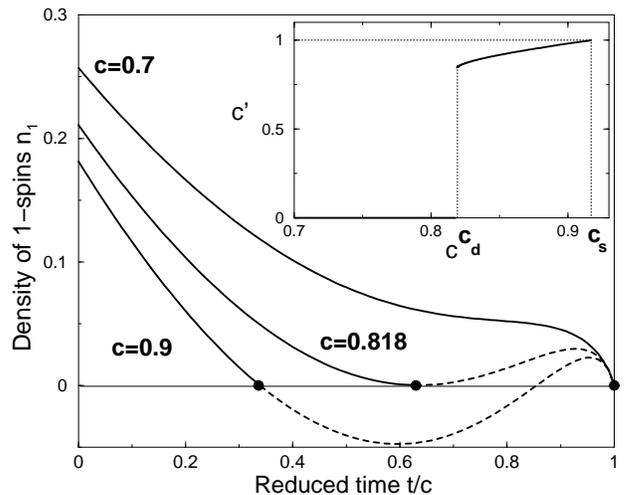}
\end{center}
\caption{Evolution of the density of 1-spins $n_1(t)$ generated by the
decimation procedure. For $c<c_d\simeq 0.818$, $n_1(t)$ remains
positive until all the plaquettes are eliminated at $t^*=c$. For
$c>c_d$ the decimation procedure stops at the time $t^*$ for which
$n_1$ vanishes (black dots), and the solution of eqn. (\ref{diff}) 
is non physical for $t>t^*$ (dashed part of the curves). 
Notice that $t^*$ discontinuously jumps down at $c=c_d$ 
(first order transition). Inset: plaquette density $c'$ for the 
reduced Hamiltonian $H'$ vs. $c$. At $c=c_d$, $c'$ discontinuously jumps to
a positive value; the threshold $c'=1$ for the disappearance of
unfrustrated GS is reached for $c_s\simeq 0.918$.}
\label{algo}
\end{figure}

\end{document}